\documentclass[12pt,prd,aps,superscriptaddress,preprintnumbers,amsmath,amssymb,nofootinbib]{revtex4-1}
\usepackage{amsmath,graphics,epsfig}
\usepackage{color,hyperref}
\usepackage{datetime}
\usepackage{booktabs}
\begin{document}

\title{Effective Higgs Theories in Supersymmetric Grand Unification}
\author{ Sibo Zheng}
\affiliation{Department of Physics, Chongqing University, Chongqing 401331, P. R. China}
\date{June, 2017}

\begin{abstract}
The effective Higgs theories at the TeV scale in supersymmetric $SU(5)$ grand unification models are systematically derived.  
Restricted to extensions on $\mathbf{5}_{H}$ containing the Higgs sector we show that only two types of real (vector-like) models and one type of chiral model are found to be consistent with perturbative grand unification.
While the chiral model has been excluded by the LHC data,
the fate of perturbative unification will be uniquely determined by the two classes of vector-like models.
\end{abstract}

\maketitle

\section{Introduction}
The Standard Model (SM)-like Higgs scalar discovered at the LHC \cite{1207.7214,1207.7235} is a milestone in the journey of exploring the nature of both electroweak symmetry breaking (EWSB) and dark mater as a weakly-interacting massive particle (WIMP). 
Firstly, the hierarchy between the established weak and Planck scale requires a novel mechanism to stabilize the radioactive correction to the Higgs mass. 
Secondly, the WIMP communicates to the SM quarks and leptons only via either the neutral $Z$ boson or Higgs scalar if no associated new particles exist. 

Five decades have passed since the idea of supersymmetry (SUSY) was firstly proposed to address the two puzzles above. For a modern review, see, e.g, \cite{Martin}.
The gauge anomaly free conditions inevitably require some amount of extension on the SUSY Higgs sector. 
For example, two Higgs doublets $H_{u}$ and $H_{d}$ are required in the minimal supersymmetric standard model (MSSM).
Since different extensions will lead to different explanations of EWSB and WIMP dark matter,
a question - how to distinguish them arises.

In this letter, we use the principle of perturbative grand unification (GUT) to systematically identify these extensions \footnote{It is not clear yet how to address non-perturbative GUT in a systematic way.},
which is one of most important motivations for SUSY.
Similar to the SM case \cite{GG}
the SUSY version of GUT can be realized through embedding the SM gauge group into a single group $SU(5)$ \cite{DG,Sakai,DRW} with rank $4$ or larger group such as  $SO(10)$ \cite{SO} and $E_6$ \cite{E6}.
For reviews, see, e.g., \cite{Langacker} and \cite{9911272}.

In what follows, we firstly consider all gauge invariant extensions on the Higgs sector that are consistent with SM gauge anomaly free conditions. See Table.\ref{Anomaly} for details.
Then we discuss which pattern survives based on the perturbative GUT.
We find that only two types of real (vector-like) models and one type of chiral model are consistent with the perturbative unification.
Since the chiral model (i.e., a fourth generation) has been excluded by the LHC data,
we conclude that the fate of perturbative unification for only extensions on $\mathbf{5}_{H}$ containing the Higgs sector  will be uniquely determined by the vector-like models.

\section{Anomaly}
The content of extra matter beyond MSSM is composed of supermultiplets under fundamental representation of $SU(5)$. 
They are constrained by the SM gauge anomaly free conditions. 
Generally it is achieved in two different ways.\\

1. The first class of construction is the so called real (vector-like) models, 
where the anomaly between each chiral supermultiplet and its conjugate is cancelled. 
This kind of interesting choices with gauge invariance are summarized in the top class in Table.\ref{Anomaly}. 
The first two models were firstly discussed in \cite{Moro, 0410085, 0807.3055}, and referred to LND and QUE in \cite{0910.2732,1006.4186} respectively.
The representations of chiral supermultiplets $Q$, $U$, $E$, $L$, $D$  under $SU(3)_{c}\times SU(2)_{L}\times U(1)_{Y}$ are given by,
\begin{eqnarray}{\label{chiral}}
\mathbf{5}&=& L(1,2,1/2)+D(3,1,-1/3),\nonumber\\ 
\mathbf{10}&=& Q(3,2,1/6)+U(\bar{3},1,-2/3)+E(1,1,1).
\end{eqnarray}
The decompositions of higher dimensional representations such as $\mathbf{15}=[1,1]$, $\mathbf{24}=[4,1]$, $\mathbf{40}=[2,1]$, etc., can be similarly derived as follows:
\begin{eqnarray}{\label{chiral2}}
\mathbf{15}&=&(6,1,-2/3)+(3,2,1/6) + (1,3,1),\nonumber\\ 
\mathbf{24}&=&(8,1,0)+(3,2,-5/6)+(\bar{3},2,5/6)+(1,3,0)+(1,1,0),\nonumber\\ 
\mathbf{40}&=&(8,1,-1)+(\bar{6},2,-1/6)+(\bar{3},3,2/3)+(\bar{3},1,2/3)+(3,2,-1/6)+(1,2,3/2)
,\nonumber\\ 
\mathbf{45}&=&L+D+(8,2,-1/2)+(\bar{6},1,1/3)+(\bar{3},2,7/6)+(\bar{3},1,-4/3)+(3,3,1/3),\nonumber\\
\mathbf{50}&=&(8,2,-1/2)+(6,1,-4/3)+(\bar{6},3,1/3)+(\bar{3},2,7/6)+(3,1,1/3)+(1,1,2)\nonumber\\
\mathbf{75}&=&(8,3,0)+(8,1,0)+(6,2,-5/6)+(\bar{6},2,5/6)+(\bar{3},2,-5/6)+(\bar{3},1,5/3)+(3,2,5/6)\nonumber\\
&+&(3,1,-5/3)+(1,1,0)
\end{eqnarray}

Note that any number of singlet chiral superfields can be added without violating the anomaly free conditions. 
Also, the combination of any two vector-like constructions in the Table such as the $\mathbf{5}+\bar{\mathbf{5}}+\mathbf{10}+\bar{\mathbf{10}}$ model \cite{Moro} is also anomaly free.\\

2. The second class of construction is chiral,
where each chiral supermultiplet introduces an anomaly, but 
the total contribution is cancelled among them \cite{ArandaWH,SlanskyYR}. 
This kind of choices with gauge invariance is outlined in the bottom class in Table.\ref{Anomaly}.
For example, the first three classes are constructed according to their anomaly coefficients $A(r)=\{1,1,6\}$
for $r=\{\mathbf{5},\mathbf{10}, \mathbf{45}\}$ respectively.
In this class the $\mathbf{5}+\bar{\mathbf{10}}$ or $\bar{\mathbf{5}}+\mathbf{10}$ model is of special interest for a $4$-th generation of supermultiplets composed of $\bar{Q}, \bar{U}, D, L, \bar{E}$ or its conjugate doesn't violate the gauge anomaly free conditions.

\begin{table}
\begin{center}
\begin{tabular}{ccc} 
$\text{Gauge Invariant Superpotential}$ ~~~~& $\text{Extra Matter}$ & ~~~$\delta \text{b}_{i}$
\\
\hline\hline
 $\mathbf{1}\cdot \mathbf{5} \cdot \bar{\mathbf{5}}_{H}+\text{H.c}$ &  $\mathbf{1}+\mathbf{5}+\bar{\mathbf{5}}$  & $1$  \\
 $\bar{\mathbf{10}}\cdot \mathbf{10} \cdot \bar{\mathbf{5}}_{H}+\text{H.c}$ &  $\mathbf{10}+\bar{\mathbf{10}}$  & $3$ \\
 $\bar{\mathbf{5}}\cdot \mathbf{15} \cdot \bar{\mathbf{5}}_{H}+\text{H.c}$ &  $\mathbf{5}+\bar{\mathbf{5}}+\mathbf{15}+\bar{\mathbf{15}}$  & $8$  \\
 $\bar{\mathbf{5}}\cdot \mathbf{24} \cdot \bar{\mathbf{5}}_{H}+\text{H.c}$ &  $\mathbf{5}+\bar{\mathbf{5}}+\mathbf{24}+\bar{\mathbf{24}}$  & $11$  \\
 $\bar{\mathbf{10}}\cdot\bar{ \mathbf{40}} \cdot \bar{\mathbf{5}}_{H}+\text{H.c}$ &  $\mathbf{10}+\bar{\mathbf{10}}+\mathbf{40}+\bar{\mathbf{40}} $  & $25$ \\
 $\bar{\mathbf{10}}\cdot \mathbf{45} \cdot \mathbf{5}_{H}+\text{H.c}$ &  $\mathbf{10}+\bar{\mathbf{10}}+\mathbf{45}+\bar{\mathbf{45}}$  & $27$ \\
 $\bar{\mathbf{24}}\cdot \bar{\mathbf{45}} \cdot \mathbf{5}_{H}$ +\text{H.c}&  $\mathbf{24}+\bar{\mathbf{24}}+\mathbf{45}+\bar{\mathbf{45}} $  & $34$ \\
 $\bar{\mathbf{40}}\cdot \mathbf{50} \cdot \mathbf{5}_{H}+\text{H.c}$ &  $\mathbf{40}+\bar{\mathbf{40}}+\mathbf{50}+\bar{\mathbf{50}} $  & $57$ \\
 $\bar{\mathbf{45}}\cdot \mathbf{75} \cdot \mathbf{5}_{H}+\text{H.c}$ &  $\mathbf{45}+\bar{\mathbf{45}}+\mathbf{75}+\bar{\mathbf{75}} $  & $74$ \\
\hline\hline
 $\mathbf{5}\cdot \bar{\mathbf{10}}\cdot \mathbf{5}_{H}$ &  $\mathbf{5}+\bar{\mathbf{10}}$  & $2$ \\
 $\bar{\mathbf{5}}\cdot \mathbf{10}\cdot \bar{\mathbf{5}}_{H}$ &  $\bar{\mathbf{5}}+\mathbf{10}$  & $2$ \\
 $\bar{\mathbf{10}}\cdot \mathbf{45} \cdot \mathbf{5}_{H}$& $\mathbf{45}+ m\cdot \bar{\mathbf{5}}+n \cdot\bar{\mathbf{10}}$ (m+n=6)  & $12+m/2+3n/2$ \\
 $\bar{\mathbf{40}}\cdot \mathbf{50} \cdot \mathbf{5}_{H}$&  $\mathbf{5}+ \bar{\mathbf{40}}+\mathbf{50}$ & $29$ \\
  \hline\hline
\end{tabular}
\caption{Gauge invariant superpotentials for different fundamental representations of $SU(5)$ which satisfy the gauge anomaly free conditions. 
The top and bottom class corresponds to real and chiral GUT models, respectively. 
Here, $\mathbf{5}_{H}$ and $\bar{\mathbf{5}}_{H}$ contains Higgs doublet $H_{u}$ and $H_{d}$, respectively.
The last column represents contribution to the coefficients of one-loop beta functions for SM gauge coupling constants.}
\label{Anomaly}
\end{center}
\end{table}

\section{Perturbative Unification}
Now we examine which type of model in Table.\ref{Anomaly} is consistent with perturbative GUT.
According to \cite{Machacek1,Machacek2,9311340} 
the one-loop renormalization group equations (RGEs) for the SM gauge couplings are given by,
\begin{eqnarray}{\label{RGE}}
\frac{d}{dt}\alpha^{-1}_{i}=-\frac{b_{i}}{2\pi},
\end{eqnarray}
where $t=\text{ln}\mu$ and 
\begin{eqnarray}{\label{b}}
b_{i}=-\{\frac{11}{3}C^{i}_{2}(G)-\frac{4}{3}\cdot\kappa\cdot T(r_{f_{i}})-\frac{1}{6}T(f_{s_{i}})\}
\end{eqnarray}
Here, $\kappa=1/2 (1)$ for two (four)-component spinor,
and $T(r)$ denotes the Dynkin index for representation $r$.
The $b_i$ coefficient is extracted from the SM gauge wave function renormalization,
which only depends on details of the representations at one-loop level.
When there are extra matter beyond MSSM, 
the beta function coefficient $b_{i}=b^{\text{MSSM}}_{i}=(33/5, 1, -3)$ will be modified by the extra matters' contribution $\delta b_{i}$ through the dynkin index $T(r)$,
the sign of which is always positive.
Note that the dynkin index of each representation in Eq.(\ref{chiral2}) depends on the details of  the representation \cite{SlanskyYR}.  
In Table.\ref{Anomaly} the value of $\delta b_{i}$ for each representation is explicitly shown in the last column.

For perturbative unification to occur, there are two different ways.\\

\begin{figure}
\includegraphics[width=0.65\textwidth]{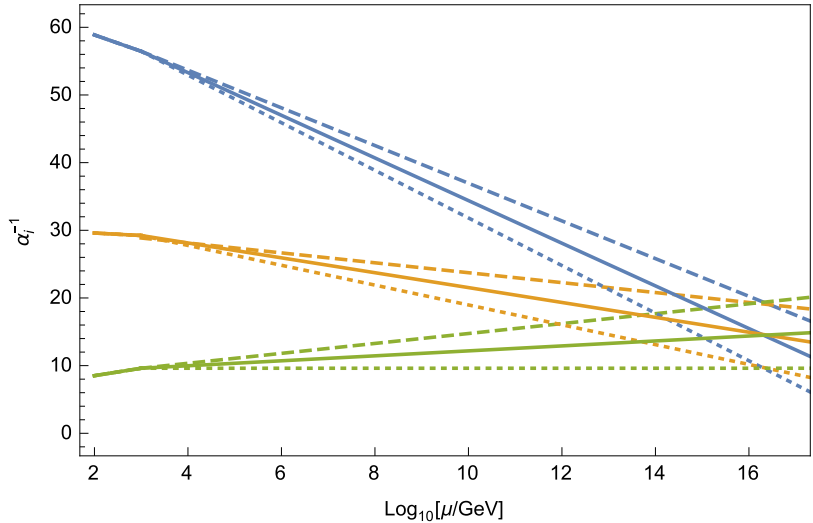}
\vspace{-0.5cm}
 \caption{One-loop RGEs for SM gauge coupling $\alpha^{-1}_{1}$ (blue), 
$\alpha^{-1}_{2}$ (orange) and  $\alpha^{-1}_{3}$ (green) in $\bar{\mathbf{5}}+\mathbf{5}$ (dashed), $\bar{\mathbf{10}}+\mathbf{10}$ (dotted) and $\bar{\mathbf{5}}(\bar{\mathbf{10}})+\mathbf{10}(\mathbf{5})$ (solid), respectively. Here we take the threshold scale $\mu=1$ TeV for illustration.}
\label{unification}
\end{figure}

1. The mass hierarchies among the extra matter are not very large, 
and unification occurs at $M_{\text{GUT}}$ before any SM gauge coupling blows up 
at smaller scale $\mu<M_{\text{GUT}}$. 
In this case only two type of vector-like models ($\mathbf{5}+\bar{\mathbf{5}}$, $\mathbf{10}+\bar{\mathbf{10}}$ and their combination $\mathbf{5}+\bar{\mathbf{5}}+\mathbf{10}+\bar{\mathbf{10}}$) 
and one type of chiral model ($\mathbf{5}+\bar{\mathbf{10}}$ or $\bar{\mathbf{5}}+\mathbf{10}$) are consistent with perturbative GUT.
Fig.\ref{unification} shows the values of $M_{\text{GUT}}$ for these GUT models for the threshold scale $\mu=1$ TeV.
In this case perturbative unification occurs in one step.\\

2. In contrast, the mass hierarchies among the extra matter are so large 
that the solution to RGEs in Eq.(\ref{RGE}) should be replaced by, e.g, for one intermediate mass scale $M_{*}>>\mu$,
 \begin{eqnarray}{\label{solution}}
\alpha^{-1}_{U}=\alpha^{-1}_{i}(M_{Z})+\frac{b_{i}-b^{\text{SM}}_{i}}{2\pi}\text{ln}\left(\frac{\mu}{M_{Z}}\right)
+\frac{b'_{i}-b_{i}}{2\pi}\text{ln}\left(\frac{M_{*}}{M_{Z}}\right)-\frac{b'_{i}}{2\pi}\text{ln}\left(\frac{M_{\text{GUT}}}{M_{Z}}\right)
\end{eqnarray}
where $b^{\text{SM}}_{i}=(41/10,-19/6,-7)$, $b_{i}$ and $b'_{i}$ represents the beta function coefficient below $\mu$, in the intermediate scale between $\mu$ and $M_{*}$, and above RG scale $M_{*}$, respectively.
In this case the appearance of $b'_{i}-b_{i}$ term \footnote{Note that $b'_{i}-b_{i}=(b^{\text{MSSM}}_{i}+\delta b'_{i})-(b^{\text{MSSM}}_{i}+\delta b_{i})=\delta b'_{i}-\delta b_{i}>0$.} in Eq.(\ref{solution}) help evade the blow up of SM gauge coupling(s) in the situation without an intermediate mass scale ($b_{i}=b'_{i}$).
With such $M_{*}$ Eq.(\ref{solution}) also shows that unification only occurs if 
\begin{eqnarray}{\label{steps}}
\delta b_{1}=\delta b_{2}=\delta b_{3}, ~~~~~ \delta b'_{1}=\delta b'_{2}=\delta b'_{3}.
\end{eqnarray}
This observation can be generalized to multiple intermediate mass scales directly.
In this case perturbative unification occurs in multiple steps.

Remarkbaly, except $\mathbf{10}+\bar{\mathbf{10}}+\mathbf{45}+\bar{\mathbf{45}}$ in the real case
and $\mathbf{5}+\bar{\mathbf{10}}$ or $\bar{\mathbf{5}}+\mathbf{10}$ in the chiral case, 
there are no such combinations for any higher dimensional representation in Eq.(\ref{chiral2}) which satisfy the condition Eq.(\ref{steps}).
As clearly shown in Eq.(\ref{chiral}), 
the success in $\mathbf{10}+\bar{\mathbf{10}}+\mathbf{45}+\bar{\mathbf{45}}$ is due to the fact that $\mathbf{45}$ contains a $\mathbf{5}$.
The effective Higgs theory at low energy scale is actually described by $\mathbf{5}+\bar{\mathbf{5}}+\mathbf{10}+\bar{\mathbf{10}}$.
For $\mathbf{5}+\bar{\mathbf{10}}$ or $\bar{\mathbf{5}}+\mathbf{10}$ with an intermediate mass scale $M_{*}$,
the extra matters beyond MSSM at the TeV scale can be either only a $\mathbf{5}$, $\bar{\mathbf{5}}$, $\mathbf{10}$ or $\bar{\mathbf{10}}$,
which corresponds to the $4$-th generation of lepton or quark supermultiplets.

Note that small deviations occur when one takes the two-loop RGEs into account, which depend on the details of both matter representations and their Yukawa interactions. 

\section{Discussions}
According to Table.\ref{Anomaly} the effective superpotential in the chiral model is given by,
\begin{eqnarray}{\label{s1}}
W_{1} = k_{d}\bar{Q} \bar{U}H_{d}+k_{u}\bar{Q}DH_{u}+k_{e} L \bar{E} H_{u},
\end{eqnarray}
and
\begin{eqnarray}{\label{s2}}
W_{2}=h_{u}QU H_{u} + h_{d}Q\bar{D}H_{d}+h_{e}\bar{L}E H_{d},
\end{eqnarray}
for $\mathbf{5}+\bar{\mathbf{10}}$ and $\bar{\mathbf{5}}+\mathbf{10}$, respectively. 
Either Eq.(\ref{s1}) or Eq.(\ref{s2}) corresponds to a fourth generation of quark and lepton supermuliplets. 
Here, the $4$-th lepton and quark masses are determined by the Yukawa coulings in Eq.(\ref{s1})- Eq.(\ref{s2}) as $k_{d}=m_{b'}/\upsilon_{u}$, $k_{u}=m_{t'}/\upsilon_{d}$, $k_{e}=m_{e'}/\upsilon_{u}$;  and $h_{d}=m_{b'}/\upsilon_{d}$, $h_{u}=m_{t'}/\upsilon_{u}$, $h_{e}=m_{e'}/\upsilon_{d}$,
where $\upsilon_{u}(\upsilon_{d})=\upsilon \sin\beta (\cos\beta)$.
Combinations of direct detections on a fourth generation of quarks at the LHC \cite{1209.0471,1209.1062,1202.5520,1202.6540} and
Higgs production cross section and decay width \cite{1111.6395,1204.1252,1111.6395} 
have excluded an explanation of perturbative fourth generation.

The two types of vector-like models may leave signatures on the following realms.
Firstly, the radiative correction to SM-like Higgs mass from the vector-like supermultiplets in Eq.(\ref{s1}) may be significant. If so, this model plays an important role in the Higgs physics.
Secondly,  the vector-like supermultiplets may give rise to significant changes in the neutralino sector,
 in which this model may play a role in WIMP dark matter.

In summary, restricted to extensions on $\mathbf{5}_{H}$ perturbative GUT delivers only two viable classes of vector-like models  ($\mathbf{5}+\bar{\mathbf{5}}$, $\mathbf{10}+\bar{\mathbf{10}}$ and their combinations) at the TeV scale,
regardless of one-step or multiple-step unification.
The fate of perturbative unification under this scenario will be uniquely determined by the footprints of these two vector-like models either in the particle collider or WIMP dark matter experiments.

\begin{acknowledgments}
This work is supported in part by the National Natural Science Foundation of China under Grant No.11405015 and 11775039.
\end{acknowledgments}

\end{document}